\documentstyle[preprint,aps]{revtex}

\begin{document}
\draft
\title
{The mobility and diffusion of a particle coupled to a Luttinger liquid}
\author{A.~H.~Castro Neto $^{\dag}$ and A.~O.~Caldeira $^{\ddag}$}
\bigskip
\address
{
$^{\dag}$
Department of Physics, University of Illinois at Urbana-Champaign\\
1110 W.Green St., Urbana, IL, 61801-3080, U.S.A.}

\address
{$^{\ddag}$ Instituto de F\'isica ``Gleb Wataghin"\\
Universidade Estadual de Campinas\\
13083-970, Campinas, SP, Brazil\\}

\maketitle

\bigskip

\begin{abstract}
We study the mobility of a particle coupled to a one dimensional
interacting fermionic system, a Luttinger liquid. We
bosonize the Luttinger liquid and find the effective interaction
between the particle and the bosonic system. We show that the
dynamics of this system is completely equivalent to the acoustic polaron
problem where the interaction has purely electronic origin.
This problem has a zero mode excitation, or soliton, in the strong
coupling limit which corresponds to the formation of a polarization
cloud due to the fermion-fermion interaction around the particle.
We obtain that, due to the scattering of the residual bosonic modes,
the soliton has a finite mobility and diffusion coefficient at finite
temperatures which depend on the fermion-fermion interaction. We show
that at low temperatures the mobility and the diffusion coefficient
are proportional to $T^{-4}$ and $T^5$ respectively and at high temperatures
the mobility vanishes as $T^{-1}$ while the diffusion increases as $T$.
\end{abstract}

\bigskip


\narrowtext

The interest on the dynamics of particles coupled to one dimensional
systems has increased lately due to the unusual properties of
these systems with respect to Fermi liquids \cite{nicolai} and their
similarity to the strongly interacting electronic systems such as
the cuprates \cite{varma}. These properties appear in the anomalous exponents
in response and correlations functions \cite{emery} and in tunneling properties
at low temperatures \cite{kane}.
All these anomalies are related with the intrinsic non-linear
character of interactions in one dimension which are due essentially
to the lack of phase space for scattering.
The topological character of the excitations of one
dimensional systems has been subject of discussion for many years and there
is an enormous amount of literature in this subject \cite{raja}.
More recently it has been found that the low energy physics of
interacting electronic
models on the lattice, such as the Hubbard chain \cite{carmelo}, or interacting
bosonic field theories in one dimension, such as the bosonic model
with local interactions \cite{bosonico}, have soliton like excitations.
Indeed, the soliton formation is one of the main characteristics of
non-linear field theories in one dimension and we have shown recently
that the soliton in a quantum field theory undergoes brownian motion
due to the scattering with environmental excitations \cite{pre}. The
same type of brownian motion can be found in the motion of ferromagnetic
domain walls in higher dimensions \cite{stamp}.

In this paper we discuss the problem of the formation of soliton like
excitations in one dimensional systems from the point of view of
an external particle which is added to the system. The only difference
between the external particle and the particles of the one dimensional
system is distinguishability, that is, we can follow the motion of the
external particle without confusing it with the environment. Then we
can make predictions for its mobility and diffusion in this environment.
Of course, a more profound problem would be to calculate the mobility
of particles inside the system without making use of the artifact of
distinguishability. Here we look at this problem from a semiclassical
point of view, that is, from the point of view of a RPA type scheme
\cite{pines}.
We want to understand the formation of a polarization cloud, and therefore
the birth of the quasiparticle, by probing the system with an external
particle. It is worth noticing that although one dimensional interacting
electronic systems, such as Luttinger liquids, are not Fermi liquids
because of the vanishing of the quasiparticle residue, their main properties
can be described within the RPA scheme \cite{fermi1}. This result is true
because the RPA fulfills all the sum rules at long wavelengths and low energies
and then it is expect to describe the continuum limit of these models.
Moreover, when indistinguishability is taken into account it is natural
to suppose that instead of an isolated soliton excitation, or quasiparticle,
we end up with a very complex non-linear theory of interacting solitons
which destroys the quasiparticle character of the excitations.

We limit ourselves to the problem of an external particle of mass $m$
moving in a lattice with intersite distance $a$ and interacting via a
local interaction potential with a system of
interacting particles with the same mass and with same potential
interaction. This problem can be thought as the problem of the Hubbard
chain with one spin down in a sea of spins up \cite{frenkel}.
By diagonalizing exactly
the fermionic system via bosonization we obtain the effective interaction
between the particle and the bosonic system, the Luttinger liquid
\cite{haldane}. By
transforming this problem back to real space we obtain a hamiltonian
which is equivalent to the problem of an electron interacting with acoustic
bosons, that is the acoustic polaron problem \cite{neto3}. It means that
the particle
is dressed by the fermions of the environment in such a way to form a soliton
like excitation. This soliton, or quasiparticle, acquires a new mass which
now depends on the interaction within the system and a renormalized coupling
constant with the bosonic environment \cite{pre}. Using the same technique
employed for the strong coupling limit of the polaron problem we calculate
the mobility and diffusion coefficients of the soliton as a function of the
fermion-fermion
interaction and temperature. This calculation is a clear example of the
non-linear character of interacting one dimensional fermionic systems and
illuminates many aspects of the formation of such type of excitations
in one dimension.

Our starting point is the hamiltonian for a particle interacting with an
interacting fermionic system (we consider spinless fermions for simplicity),
\begin{equation}
H = H_P + H_I + H_F,
\end{equation}
where $H_P$ is the hamiltonian for the particle alone,
\begin{equation}
H_P = \frac{p^2}{2 m},
\end{equation}
$H_I$ is given by,
\begin{equation}
H_I = \sum_{j=1}^{N} U(x-x_j),
\end{equation}
and it couples the particle to a set of $N$ other
particles at positions $x_j$ ($j=1,2,3,...,N$).
The fermionic system is described by,
\begin{equation}
H_F = \sum_{j=1}^{N} \frac{p_j^2}{2 m} +
\frac{1}{2} \sum_{i,j=1; j \neq i}^{N} U(x_i-x_j).
\end{equation}
In second quantized form the interaction term in (3) and (4)
can be written as $U \sum_{i} n_i n_{i+1}$ where $i$ labels the sites on
a chain.

Our first step is to diagonalize the hamiltonian (4) using the
techniques of bosonization of one dimensional systems \cite{emery}.
We will skip the details here since this technique is already well
known. It is possible to show that the hamiltonian for the fermionic
system can be written as (in our units $\hbar=K_B=1$),
\begin{equation}
H_F = \sum_{q>0} E_q \beta_q^{\dag} \beta_q,
\end{equation}
which has exactly the same form as for a set of independent harmonic
oscillators with energies given by,
\begin{equation}
E_q = v_F |q| \sqrt{1 + \frac{U}{2 \pi v_F}}.
\end{equation}
Notice that the only effect of the interaction in the spectrum is
a renormalization of the Fermi velocity, $v_F$, to a new value,
$\tilde{v}_F = v_F \sqrt{1 + \frac{U}{2 \pi v_F}}$.

In this new representation the interaction term (3) is written as,
\begin{equation}
H_I =  - i \sum_{q} \left(\frac{|q|}{2 \pi L}\right)^{1/2}
U e^{i q x -\phi} sgn(q) \left(\beta_{-q} + \beta_q^{\dag}\right),
\end{equation}
where,
\begin{equation}
\tanh(2 \phi)=\frac{U}{4 \pi v_F + U}.
\end{equation}

Now we define new operators for the bosons in the coordinate and momentum form,
\begin{eqnarray}
\beta_q = \sqrt{\frac{E_q m}{2}} \left(X_q + i \frac{P_{-q}}{E_q m}\right)
\nonumber
\\
\beta_q^{\dag} = \sqrt{\frac{E_q m}{2}} \left(X_{-q} - i \frac{P_q}{E_q
m}\right),
\end{eqnarray}
which obey $\left[X_q,P_{q'}\right] = i \delta_{q,q'}$.

The hamiltonian for the problem can also be rewritten in second quantized
form as,
\begin{equation}
H = \sum_k \epsilon_k d_k^{\dag} d_k + \sum_q \left(
\frac{P_q P_{-q}}{2 m} + \frac{m E_q^2}{2} X_q X_{-q}\right)
-i U e^{-\phi} \left(\frac{\tilde{v}_F}{\pi L}\right)^{1/2}
\sum_q q \rho(q) X_{-q}
\end{equation}
where $d_k$  and $d_k^{\dag}$ are the creation and annihilation operators
for the particle in the state of momentum $k$,
$\epsilon_k = \frac{k^2}{2 m}$ is the dispersion relation for
the particle (we have the constraint that there is just one
particle in the system, $\sum_k d_{k}^{\dag} d_k = 1$, for all states of the
Hilbert space) and $\rho(q) = \sum_k d_{k+q}^{\dag} d_k$ is the density
operator for the particle (and since the particle is a fermion,
$\{d_k,d_{k'}^{\dag}\} = \delta_{k,k'}$).

Now we go back to real space by defining the following field operators,
\begin{eqnarray}
\eta(x) = \frac{1}{\sqrt{N}} \sum_q e^{i q x} X_q;
\nonumber
\\
\Pi(x) = \frac{1}{\sqrt{L a} } \sum_q e^{-i q x} P_q;
\nonumber
\\
\psi(x) = \frac{1}{\sqrt{L}} \sum_k e^{i k x} d_k,
\end{eqnarray}
where $L = Na$ is the length of the system, $N$ is the number of sites.
It is easy to prove that the commutation
relation between these above defined operators is
$\left[\eta(x),\Pi(x')\right] = i \delta(x-x')$ and
$\left\{\psi(x),\psi^{\dag}(x')\right\} = \delta(x-x')$.

Finally we rewrite the hamiltonian as,
\begin{equation}
H = \int{dx} \left\{\frac{\Pi^2}{2 \nu} + \frac{\nu \tilde{v}_F^2}{2}
\left(\frac{\partial \eta}{\partial x}\right)^2
+ \frac{1}{2m}\frac{\partial \psi^{\dag}}{\partial x} \frac{\partial
\psi}{\partial x} + D \frac{\partial \eta}{\partial x}
\psi^{\dag} \psi\right\},
\end{equation}
where $\nu = m/a$ and
\begin{equation}
D = U e^{-\phi} \left(\frac{\nu \tilde{v}_F}{\pi}\right)^{1/2} =
U \left(\frac{\nu v_F}{\pi}\right)^{1/2}
\end{equation}
is the coupling between the fields. It is amazing to notice that
the hamiltonian (12) has the same form the hamiltonian for an
electron coupled to acoustic phonons where $D$ is the deformation
potential coupling \cite{neto3}. We could therefore apply the results
for the strong coupling limit obtained in ref. \cite{neto3} directly to
the problem of the Luttinger liquid. The strong coupling is obtained
when the deformation energy is much larger than the characteristic bosonic
energy scale, that is,
\begin{equation}
\frac{D a}{\tilde{v}_F} = \frac{u}{\sqrt{1+u}} \sqrt{4 \pi p_F a} >> 1,
\end{equation}
where $u = \frac{U}{2 \pi v_F}$ is the dimensionless coupling constant and
$p_F = m v_F$ is the Fermi momentum.

As we have shown in ref. \cite{neto3} the physics of the problem can
be understood in terms of the formation of a self consistent potential
around the particle due to the formation of the polarization cloud.
In the strong coupling limit the particle stays in the ground state
of this potential with ground state energy $E_0 = \frac{g^2}{8 m}$
and can only undergo virtual transitions to excited
states. Moreover, the deformation of the bosonic system has the form,
$\eta_0(x- v t) = -\frac{g}{2mD}\: \tanh\left(\frac{g (x- v t)}{2}\right)$
where $v$ is the soliton velocity and $g$ is the renormalized coupling
constant of the theory which is given by,
\begin{equation}
g = m \nu \left(\frac{D}{ \tilde{v}_F \nu}\right)^2
 = 4 \pi \frac{u^2}{1+u} p_F.
\end{equation}
This result shows that the displacement
of the bosonic field has topological
character, that is, it interpolates between two different vacua in
the asymptotic limit ($Lim_{x \to \pm \infty} \eta_o(x) = \mp \frac{g}{2mD}$).
Thus we see that the soliton surf-rides on the wave produced by
the distortions of the bosonic field.

The reader can find the details of the calculation in ref. \cite{neto3}.
Here we just sketch the main steps leading to our results.
We use the saddle point approximation for the static configuration in order
to generate an expansion of the form $\eta(x,t)=\eta_0(x) + \rho(x,t)$.
Then we expand the field configuration up to second order in $\rho(x,t)$
and end up with a theory for the renormalized bosons which has a zero
mode with the form,
\begin{equation}
\rho_0(x) = \sqrt{\frac{3ag}{8}} sech^2 \left(\frac{gx}{2}\right).
\end{equation}
We can also show that the soliton mass is given by \cite{neto3},
\begin{equation}
M_s = \frac{32}{3} m \left(\frac{E_o}{\tilde{v}_F g}\right)^2=
\frac{8}{3} \frac{u^4}{\left(1+u\right)^3} m.
\end{equation}
These results give the complete picture of the quasiparticle formation:
the interaction between the particle leads to the creation of a polarization
cloud which renormalizes the mass of the particle and the interaction with
the environment.

The existence of the zero mode in the problem leads naturally to the
quantization of its motion via the canonical coordinate formalism \cite{raja}.
It is possible to show that the soliton is scattered by the residual
(long wavelength) bosonic excitations in the system and it leads to
finite mobility and diffusion \cite{pre,prl}. The mobility of the soliton,
$\mu$, as a function of the temperature, $T$, is given by \cite{neto3},
\begin{equation}
\mu(T) = \frac{ 32 \pi M_s g^2}{g^2 I\left(\frac{T_c}{T}\right)}
\end{equation}
where ($E_F=v_F p_F$ is the Fermi energy),
\begin{equation}
T_c = \frac{g \tilde{v}_F}{2 k_B} =
4 \pi \frac{u^2}{\sqrt{1+u}} E_F
\end{equation}
is the characteristic temperature for the bosonic excitations and,
\begin{equation}
I(S) = S \int^{\infty}_0{d\kappa} \; \kappa^2 \; R(\kappa) \;
\frac{e^{S\kappa}}
   {(e^{S\kappa}-1)^2}
\end{equation}
depends on the reflection coefficient, $R$, of residual bosons which
are scattered by the soliton. At low temperatures, $T \ll T_c$,
we have \cite{neto3},
\begin{equation}
\mu(T)= \frac{16 \pi M_s}{27 g^2} \left(
 \frac{T_c}{T} \right)^4 \approx 8.05  \frac{u^8}{\left(1+u\right)^3}
\frac{E_F^3}{T^4}.
\end{equation}
This result shows that the soliton is almost free at low temperatures
due to the absence of bosonic degrees of freedom. Remarkably enough
in the strong coupling limit the mobility increases with the interaction
strength.
This unusual behavior is typical of the non-linear character
of this system and it has many interesting properties in the study of
tunneling in one dimension \cite{kane}.

Using the same methods we can also shown that the diffusion coefficient
in momentum space is given by,
\begin{equation}
\overline{D}(T) \approx \frac{27 g^2 \hbar}{16 \pi} \frac{T^5}{T^4_c} \approx
0.54 \frac{1}{u^4} \frac{T^5}{v_F^4 p_F^2}
\end{equation}
Again, observe that at low temperatures the soliton moves ballistically.
Observe that, contrary to the mobility, the diffusion decreases with the
increasing interaction.

At high temperatures, $T_c \ll T$, we find \cite{neto3},
\begin{equation}
\mu(T) = \frac{64 \pi^4 M_s}{315 g^2} \frac{T_c}{T} \approx 41.45
\frac{u^2}{(1+u)^{3/2}} \frac{1}{T}
\end{equation}
and
\begin{equation}
D(T) = \frac{315 g^2 T}{16 \pi^5} \approx 10.16
\frac{u^4}{\left(1+u\right)^2} p_F^2 T.
\end{equation}
Observe that both the mobility and the diffusion coefficient increase
with the interaction at high temperatures.

In summary, in this paper we calculate the mobility and diffusion
coefficients of a particle coupled to a Luttinger liquid as functions of
temperature and the coupling constant by mapping this problem in the strong
coupling regime of the problem of the acoustic polaron.We draw a picture for
the formation of a soliton like excitation
in this problem which resembles the birth of a quasiparticle. We
calculate explicitly the renormalized mass and coupling constant
as a function of the fermion-fermion interaction. This problem has
relevance for the understanding of the structure of excitations of
one dimensional interacting systems.

\acknowledgments
AHCN would like to thank D.~K.~Campbell, E.~H.~Fradkin, D.~M.~Frenkel,
A.~J.~Leggett, N.~V.~Prokof'ev and P.~C.~E.~Stamp for many
illuminating discussions and comments,
Conselho Nacional de Desenvolvimento Cient\'ifico e Tecnol\'ogico,
CNPq (Brazil), for financial support.
A.~O.~Caldeira also acknowledges CNPq for partial support.

\end{document}